\documentclass[singlespacing,floatfix,twocolumn,article,english]{revtex4}
\usepackage[english]{babel}
\usepackage{color}
\usepackage{amsmath}
\usepackage{graphicx}
\usepackage[caption=false]{subfig}
\usepackage{xspace}

\newcommand{\fb}{\ensuremath{\tilde{\Delta G}}}
\newcommand{\fbt}{\ensuremath{\tilde{\Delta G}_\mathrm{T}}}
\newcommand{\fbu}{\ensuremath{\tilde{\Delta G}_\mathrm{U}}}
\newcommand{\fbin}{\ensuremath{\tilde{\Delta G}_\mathrm{P}}}
\newcommand{\fx}{\ensuremath{\tilde{\Delta G}_\mathrm{X}}}

\newcommand{\gammat}{\ensuremath{\gamma_\mathrm{T}}}
\newcommand{\gammaun}{\ensuremath{\gamma_\mathrm{U}}}

\newcommand{\mvf}{Martinez-Veracoecha and Frenkel }

\newcommand{\nl}{\ensuremath{N_\mathrm{L} }}
\newcommand{\nun}{\ensuremath{N_\mathrm{U} }}
\newcommand{\nt}{\ensuremath{N_\mathrm{T} }}

\makeatletter
\def\maxwidth{%
  \ifdim\Gin@nat@width>\linewidth
    \linewidth
  \else
    \Gin@nat@width
  \fi
}
\makeatother


\date{\today}

\begin{document}

\captionsetup[figure]{justification=raggedright}
\title{Exploiting Receptors Competition to Enhance Nanoparticles Binding Selectivity}
\author{Stefano Angioletti-Uberti$^{1,2}$}
\email{sangiole@imperial.ac.uk}
\affiliation{$^1$ Beijing Advanced Innovation Centre for Soft Matter Science and Engineering, Beijing University of Chemical Technology, Beijing, PR China}

\affiliation{$^2$ Imperial College London, Department of Materials, London, UK}

\begin{abstract}
Nanoparticles functionalized with multiple ligands can be programmed to bind biological targets 
depending on the receptors they express, providing a general mechanism exploited in various technologies, 
from selective drug-delivery to biosensing. 
For binding to be highly selective, ligands should exclusively 
interact with specific targeted receptors, because formation of bonds with other, untargeted ones would 
lead to non-specific binding and potentially harmful behaviour. 
This poses a particular problem for multivalent nanoparticles, because even very weak bonds can 
collectively lead to strong binding. A statistical mechanical model is used here to describe 
how competition between different receptors together with multivalent effects can be harnessed to design 
ligand-functionalized nanoparticles insensitive to the presence of untargeted receptors, 
preventing non-specific binding.
\end{abstract}
\maketitle

Cells typically present a large variety of receptors on their surface, providing a ``biological barcode'' to distinguish cells of different type or in different states, e.g. healthy vs sick \cite{collins}. 
This idea is at the basis of a widely used drug-delivery strategy in nanomedicine, which exploits nanoparticles functionalized with multiple ligands that recognize specific receptors, whose expression is known to be associated to a disease \cite{targeted-delivery,carlson-kiessling}. In the same way, various nanoparticles-based biosensing applications also exploit ligand-receptor binding to detect the presence of specific targets \cite{review-translational}.\\
For selective targeting based on ligand-receptor recognition, strategies are required to distinguish targets based not only on the type of receptors present, but also on their expression level. The latter is particularly important for those diseases, including certain cancers, where healthy and sick cells do not present different receptors, but rather over-regulate one (or few) of them \cite{targeted-delivery}. For both targeting scenarios it has been shown that nanoparticles displaying multiple binding ligands can better discriminate targets compared to monovalent drugs \cite{cochran-1,cochran-2,carlson-kiessling,fran-pnas}. However, this is not always the case, and understanding the conditions leading to enhanced selectivity, or loss of it, is an important step for rational design. 
\begin{figure}[h]
\includegraphics[width=0.45\columnwidth]{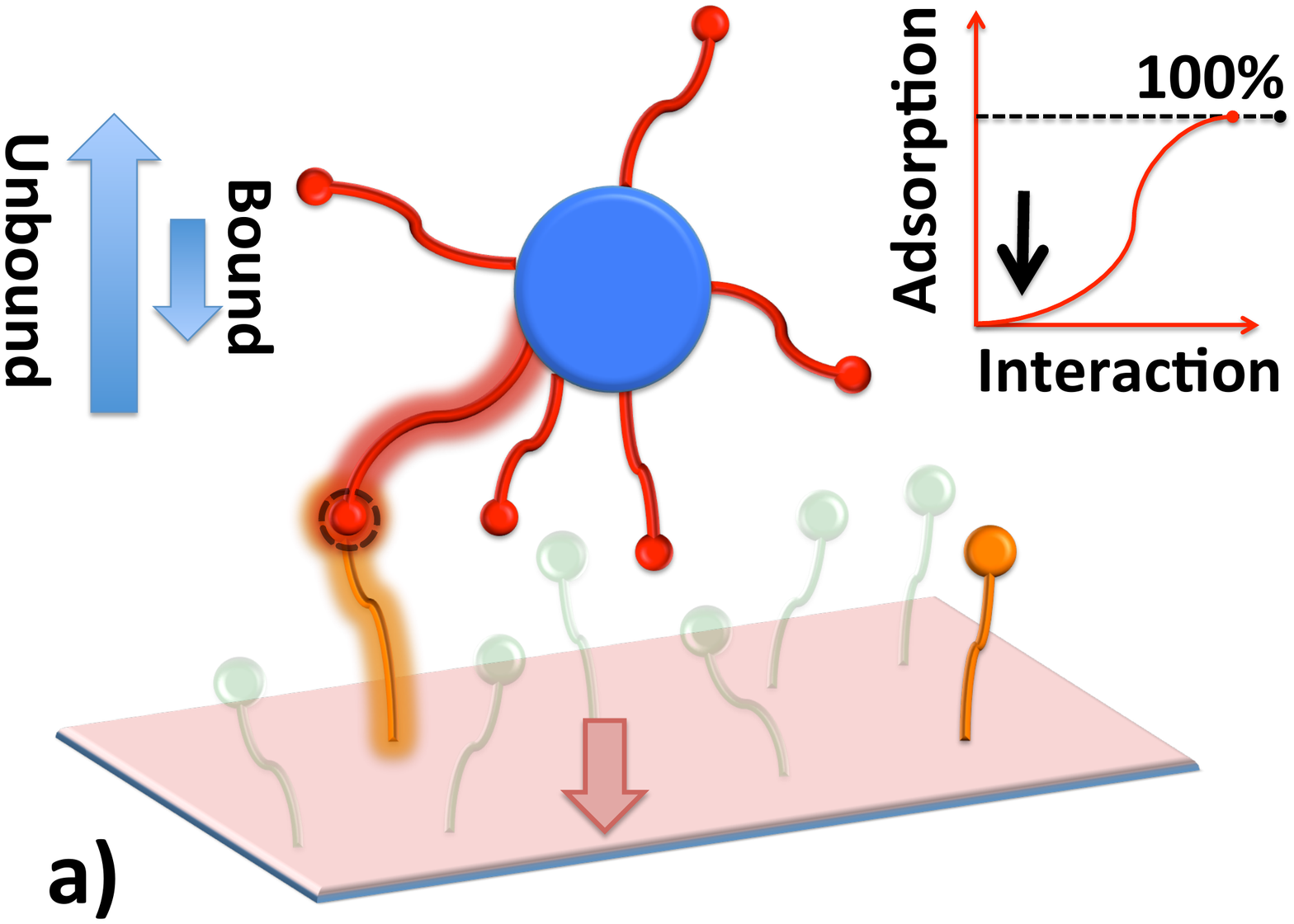}
\includegraphics[width=0.45\columnwidth]{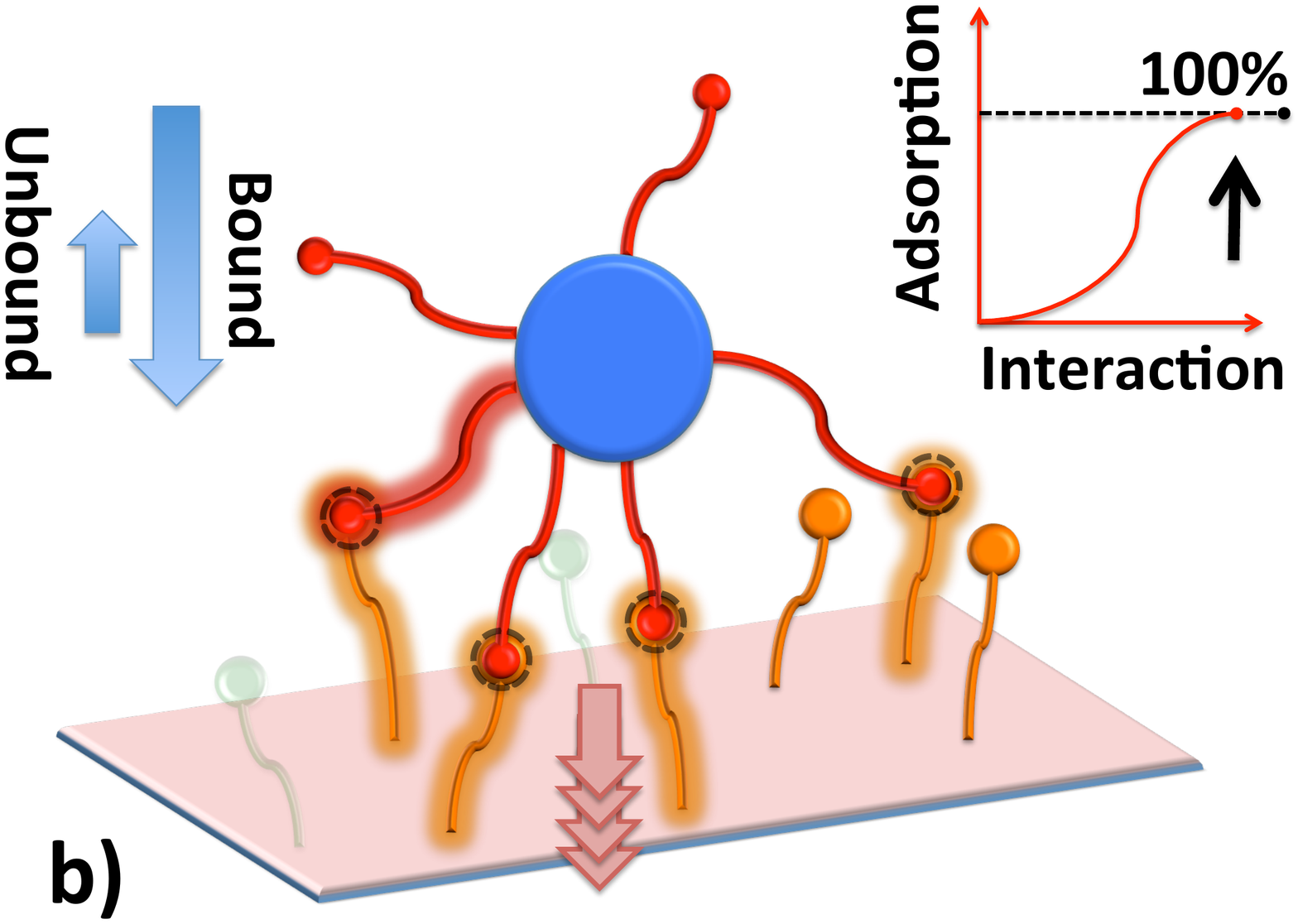}\\
\includegraphics[width=0.45\columnwidth]{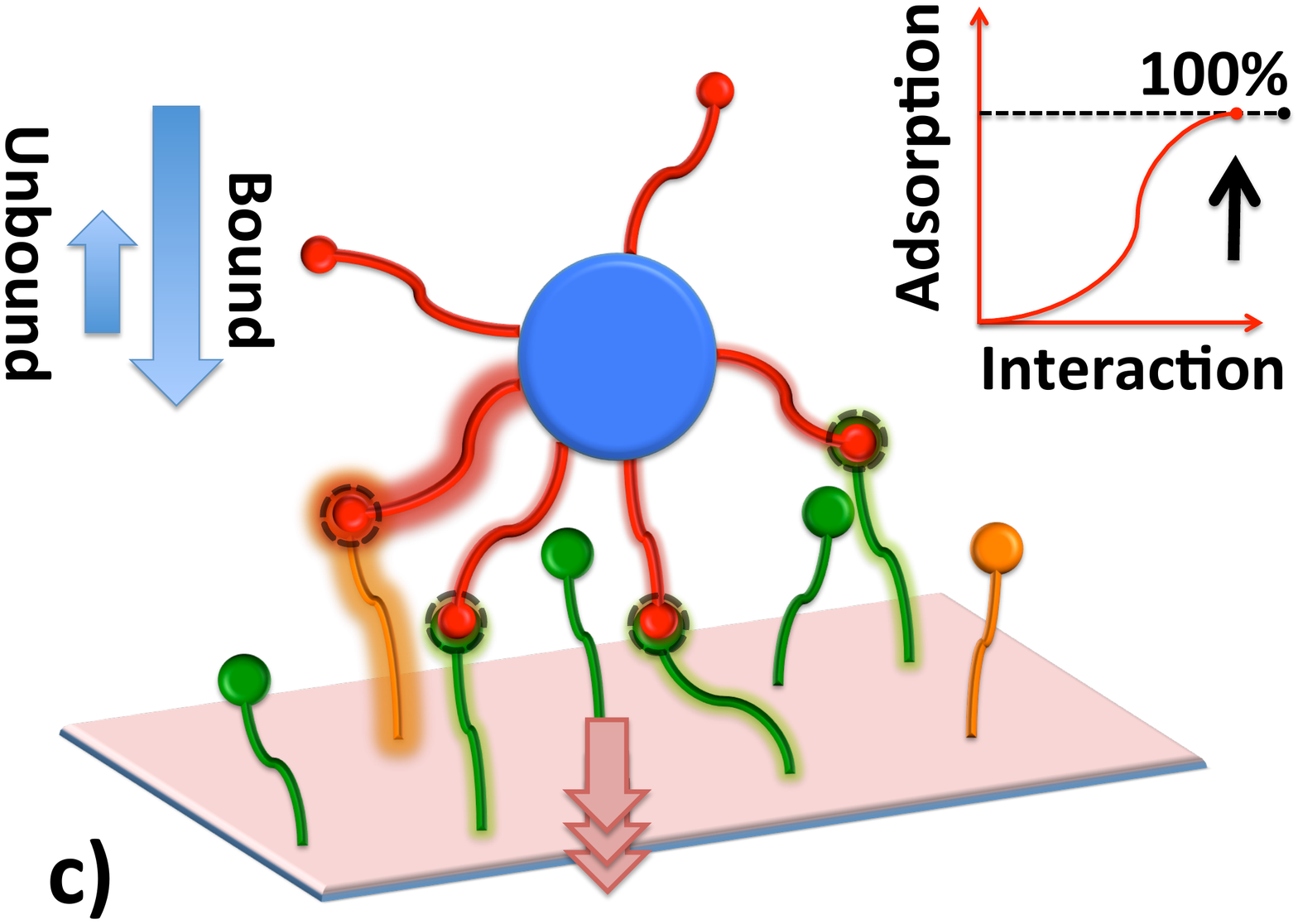}
\includegraphics[width=0.45\columnwidth]{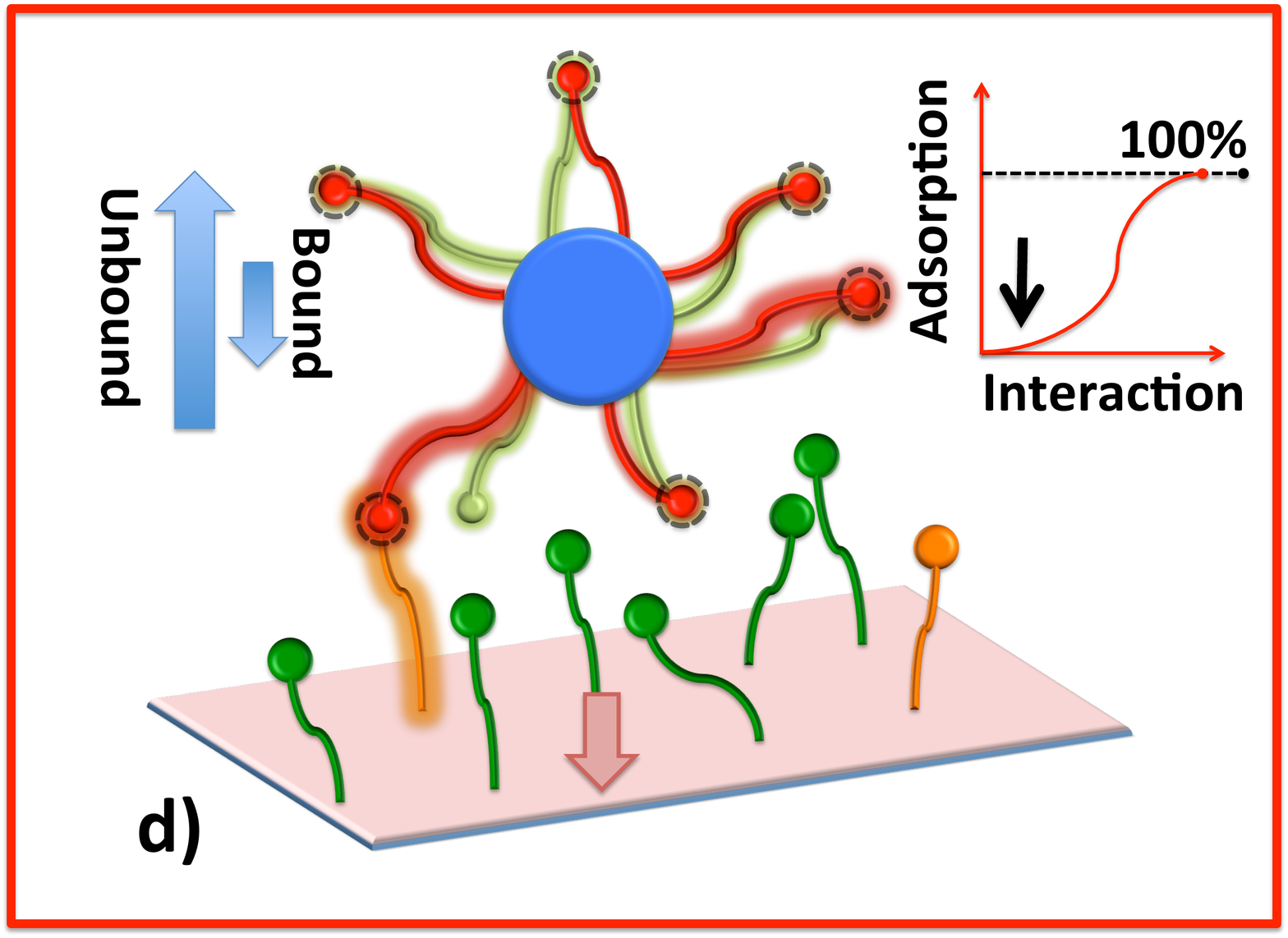}
\vspace{-.5cm}
\caption{Binding of a multivalent nanoparticle (blue sphere, ligands in red) to a cell surface (pink) expressing certain receptors. Ligands specifically target orange receptors,  (over-)expressed in a disease state. Green receptors are other receptors not related to it.
a)-b) ideal scenario: Ligands only bind to targeted receptors, and nanoparticles will adsorb when these are expressed above a certain threshold (a), but not below (b) \cite{fran-pnas}. 
c,d) realistic scenario: Ligands see both targeted and untargeted receptors, binding the latter weakly. Multiple 
weak bonds can still lead to non-specific adsorption even when targeted receptors are not over-expressed (c). 
This problem can be alleviated using ``protective'' receptors (light green) directly coated on the nanoparticle (d). If these form stronger bonds compared to untargeted receptors, selective binding is restored.}
\label{fig:cases}
\vspace{-.3cm}
\end{figure}
The mechanism behind enhanced selectivity of multivalent nanoparticles towards receptor over-expression, also dubbed ``super-selectivity'', was first elucidated by \mvf \cite{fran-pnas}. These authors used both analytical theory and Monte Carlo simulations to show how such effect could be well understood considering the statistical mechanics of multivalent binding. 
Later on, Dubacheva {\it et al} \cite{tine-jacs,tine-pnas} extended their theory to the case of targeting using multivalent polymers. In these latter works, extensive comparison to experimental results proved how such theory could correctly describe various trends observed, and thus be used to design targeting applications \cite{tine-pnas}. 
%
%
%
Previous works on the super-selectivity of multivalent nanoparticles focused on model systems, where only a single type of receptor is considered. Cells, however, express various types, most of which associated to normal function. This observation naturally leads to the question: Can these other receptors affect targeting selectivity (see Fig.~\ref{fig:cases} for reference)? Due to the physics of their interaction, this is particularly important for multivalent constructs.
Consider for example a nanoparticle whose ligands are optimized to target one specific receptor type. To be more precise, this means that the strength of a bond with such receptor is higher compared to any other. The problem lies in the fact that due to multivalency weak single bonds to untargeted receptors can still collectively provide a high binding energy \cite{whitesides}. This in turn can drive non-specific adsorption of the nanoparticles to cells which do not express the targeted receptors. In the context of drug-delivery, this scenario would lead to higher toxicity, whereas in biosensing it would cause false positives.
%
%
%
%
%
In this Letter, we combine a recent theory of ligand-receptor-mediated interactions \cite{stefano-jcp,patrick-jcp} 
with the model developed by \mvf for nanoparticles adsorption (details in the SI) to quantify this scenario, and propose 
a general solution.\\
Briefly, the \mvf model describes adsorption of ligand-functionalized nanoparticles from a solution to a target displaying multiple receptors. Expanding this model to account for multiple receptor types, i.e. targeted and untargeted, the adsorption probability $\theta$ reads:
\begin{equation}
\theta = <\frac{ z q\left( \nl, \nt, \nun, \{\fb_{\mathrm{X}}\} \right)}{1 + z q\left( \nl, \nt, \nun, \{\fb_{\mathrm{X}}\} \right)}>_{<N_T>,<N_U>}.
\label{eq:adsorption}
\end{equation}

Subscripts $L,T,U$ refer to ligands, targeted and untargeted receptors, respectively. Hence in Eq.\,\ref{eq:adsorption} $N_{X=L,T,U}$ is the number of ligands and targeted or untargeted receptors, whereas $\fb_\mathrm{X}$ is the bond free energy $\fb$ of a ligand with receptor of type $X = T,U$ (we always consider energies to be scaled by the thermal energy $k_\mathrm{B} T$, where $k_\mathrm{B}$ is Boltzman's constant and $T$ temperature). $<>_{<N_T>,<N_U>}$ indicates an average over two Poisson distributions, one over the number of targeted receptors, with average $<\nt>$, and one over the number of untargeted receptors, with average $<\nun>$. Furthermore, $q$ is the ratio between the partition function of the nanoparticle in the bound vs unbound state and $z$ is the nanoparticles' activity in the bulk solution, proportional to nanoparticles' number density and binding volume.  For ease of discussion, we focus here on the case of only two receptor types. However, it can be shown that any arbitrary distribution of untargeted receptors can be mapped onto, or be bound by, a two-types model with an effective number of receptors. For this reason, results do not change upon considering more complex scenarios (see the SI).\\
Eq.\,\ref{eq:adsorption} has the form of the Langmuir adsorption isotherm. All multivalent effects enter via $q$, which can be written as:
\begin{equation}
q = e^{-\left(\tilde{F}_{\mathrm{att}}^{\mathrm{surf}}-\tilde{F}_{\mathrm{att}}^{\mathrm{bulk}} \right)} - 1,
\label{eq:fads}
\end{equation}
where $\tilde{F}_{\mathrm{att}}$ indicates the free-energy due to bond 
formation, and $\tilde{F}_{\mathrm{att}}^{\mathrm{surf}}$, $\tilde{F}_{\mathrm{att}}^{\mathrm{bulk}}$ are its values when the nanoparticle is either adsorbed on the target surface or in the bulk solution, respectively. A formal definition of $\tilde{F}_{\mathrm{att}}$, which is always negative or at most zero when no bonds can form, is given in \cite{stefano-review}. The $-1$ term in Eq.\,\ref{eq:fads} takes into account the constraint that a particle must have at least one bond with the surface to be bound. 
Note that $\tilde{F}^\mathrm{bulk}_\mathrm{att}$ is zero if nanoparticles can only form bonds with the surface. 
However, this is not true when {\it intra-particle} bonds occur, an important fact exploited later.
Given Eqs.~\ref{eq:adsorption},\ref{eq:fads}, the adsorption probability $\theta$ can be calculated once $\tilde{F}_{\mathrm{att}}$ is known for the system of ligand-receptor pairs considered. 
This is provided by the formula \cite{stefano-jcp}:

\begin{equation}
\tilde{F}_{\mathrm{att}} = \sum_{i} \log p_i + \frac{1}{2} \left( 1 - p_i \right)
\label{eq:ste1}
\end{equation}

where $i$ is an index running over all possible binders, i.e. ligands {\it or} receptors, and $p_i$ is 
the probability that such binder is unbound, given by the 
solution of a set of non-linear coupled equations:

\begin{equation}
p_i + \sum_j p_i p_j \chi_{ij} = 1,
\label{eq:ste2}
\end{equation}

where $\chi_{ij} = \exp\left(  -\fb_{ij} \right)$. In Eq.\ref{eq:ste2}, $\chi_{ij}$ is the strength of a bond for a specific ligand-receptor pair $i,j$ and $\fb_{ij}$ is the corresponding bond energy. 
The index $j$ runs over all possible binding partners of $i$, and there are $N_{\mathrm{binder}}$ coupled equations, one for each binder in the system. This set of equations must be solved numerically \cite{patrick-jcp}, but analytical expressions from a mean-field model providing a qualitative picture will also be discussed (see the SI).\\
%
%
%
We illustrate our results by calculating two key quantities describing binding selectivity. 
The first is the adsorption probability $\theta$, given by Eq.\,\ref{eq:adsorption}. The second is the so-called selectivity parameter $\alpha$ \cite{fran-pnas}, defined here as:

\begin{equation}
\alpha( <N_T>, <N_U> ) = \frac{\partial\log \theta}{\partial\log <N_{\mathrm{T}}>}|_{<N_T>,<N_U>}
\label{eq:alpha}
\end{equation}

which gauges nanoparticles' ability to tell apart binding sites with different numbers of targeted receptors $<N_{\mathrm{T}}>$. 
More precisely, for $\alpha>1$ the adsorption probability raises super-linearly, approaching an ideal on-off behaviour where particles bind exclusively if receptors concentration is higher than a certain threshold. On the opposite side, $\alpha<1$ indicates appreciable adsorption for a broad range of receptors concentrations, $\alpha=0$ being indiscriminate adsorption.
%
\begin{figure}[h!]

\includegraphics[width=.5\textwidth]{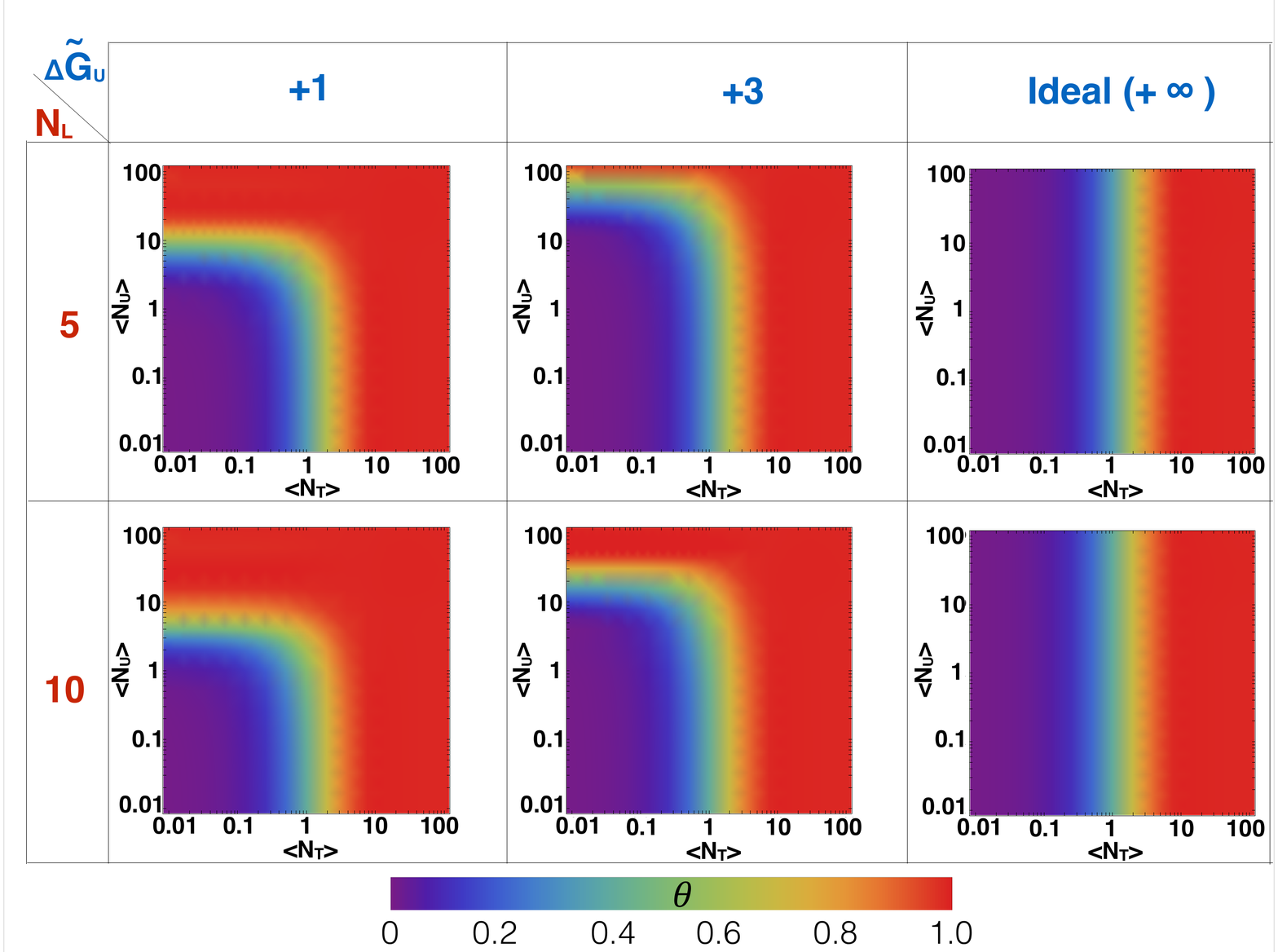}
\includegraphics[width=.53\textwidth]{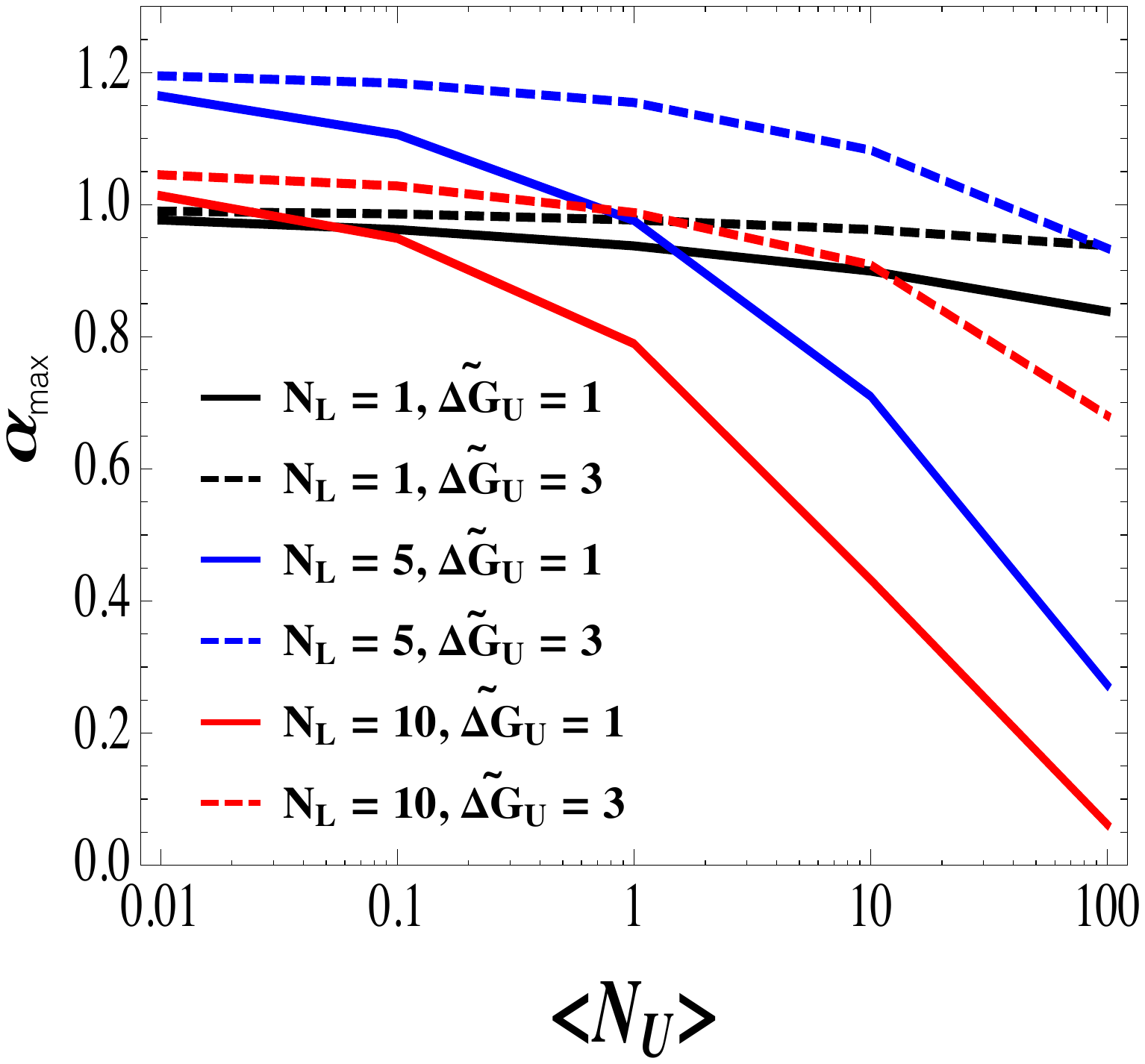}
\vspace{-1.5cm}
\caption{Top: Adsorption probability $\theta$ ( Eq.\,\ref{eq:adsorption}) as a function of the number of targeted ($\nt$) and untargeted ($\nun$) receptors. Results are reported for different number of ligands $\nl$ and untargeted bonds energies $\fbu$ (the energy of a targeted bond is $\fbt=-3$). As the number of untargeted receptors increases, their effect becomes stronger and nanoparticles adsorb also when in the ideal case (rightmost panels), this would not occur. This problem increases with higher valencies and bond strengths (smaller $\fbu$). Bottom: Maximum selectivity $\alpha$ achievable, Eq.\,\ref{eq:alpha}, as a function of $\nun$.}
\label{fig:real-binding}
\end{figure}

Fig\,\ref{fig:real-binding} illustrates the effect of non-specific interactions for nanoparticles differing in either the number of targeting ligands $\nl$ or the bond strength with untargeted receptors. The rightmost panels in the top figure represent the ``ideal'' behaviour one would observe when ligands do not interact at all with untargeted receptors (see Fig.\,\ref{fig:cases}$a$). In this case, appreciable adsorption is observed only above a certain number of targeted receptors. 
When the strength of untargeted bonds increases, the effect of non-specific interactions rapidly increases too and particles also adsorb when this is not expected. In fact, above a certain number of untargeted receptors appreciable adsorption will occur even when the targeted receptors are completely absent, completely losing selectivity. 
As can be inferred by comparing nanoparticles with different number of ligands, this problem becomes more prominent as their valency increases. The observed behaviours can be qualitatively captured within a mean-field description, which provides the simple expression: 
\begin{equation}
q = \left( 1 + \gammat + \gammaun \right)^{\nl}-1; \quad \gamma_{\mathrm{x}} = \\
N_{\mathrm{x}} \exp\left(-\fx\right).
\label{eq:mean-field-classic}
\end{equation}
%
Given the dependence of $\theta$ on $q$ (Eq.~\ref{eq:adsorption}) appreciable adsorption is expected whenever $q \gtrsim z^{-1}$. From Eq.~\ref{eq:mean-field-classic}, $q$ increases with both the strength of untargeted receptors and the number of ligands \nl. Notably, this increase is super-linear with respect to the number of untargeted receptors whenever the valency is higher than one, predicting that selectivity will rapidly deteriorate for multivalent nanoparticles once interactions with untargeted receptors are accounted for. The same expression also shows that whenever the number of untargeted receptor is higher than an upper limit $\nun^\mathrm{lim} = \exp(\beta\Delta G_\mathrm{U} ) \left[ (1/z+1)^{1 / \nl} - 1\right]$, adsorption occurs even in the complete absence of {\it any} targeted receptor.\\
In most applications, multivalency have been exploited to increase the binding strength to a specific target using singularly weak-bonds. For this use, the binding strength enhancement has the positive effect to decrease the detection level of the target. However, what we show here is that exactly this same feature can provide enough binding strength to drive adsorption to the wrong targets due to the formation of many weak bonds with untargeted receptors. 
Hence, when considering targeting applications, care must be taken with multivalent architectures.\\  
%
%
The analysis of the selectivity parameter $\alpha$, reported in Fig\,\ref{fig:real-binding}, provides further insights into the extent of this problem. More precisely, for each value of $<\nun>$ we consider the maximum $\alpha$ achievable within the $<\nt>$ range, which provides an upper bound to the selectivity of the system.
When only targeted receptors are present, it was shown \cite{fran-pnas} that multivalent nanoparticles can achieve almost optimal on-off behaviour, i.e. $\alpha>1$ around some value of $\nt$. In practice, this means that nanoparticles adsorb only when the number of receptors is higher that $\nt$, and never below. This so-dubbed ``super-selectivity'' is a specific feature of multivalent particles that, importantly, monovalent ones can never achieve \cite{fran-pnas,tine-pnas,tine-jacs}. It is this advantage that suggest the use of multivalent nanoparticles for selective targeting, considering that in various diseases cells simply over-regulate the expression of certain receptors rather than expressing a mutated form that could be distinguished using an optimized ligand. 
Considering the effect of untargeted receptors, however, the picture changes. As shown in Fig\,\ref{fig:real-binding}, the super-selective regime can completely disappear, and indeed under broad conditions multivalent particles will perform worse then monovalent ones. As for the case of $\theta$, the observed trends are a decrease in selectivity with increasing valency and increasing strength of the bonds to untargeted receptors, as should be expected.\\
Given the problems previously discussed and the trends observed, a question arises: Can we simply optimize ligands to reduce non-specific interactions? The answer to this question is indeed positive, but one must solve a complex optimization problem. Ligands optimization is usually run to find the one with the highest possible binding strength for a given receptor \cite{review-translational}. This is a perfectly valid strategy if the aim is to decrease their detection threshold. This strategy, however, does not necessarily improve the ability of ligands to discern between different receptor types, which requires increasing the gap in the strength between the targeted receptor and {\it every other possible receptor type}, a much harder optimisation problem. Additionally, although increasing the strength of the bond with targeted receptors improve their detection
, it also negatively impacts the possibility to tell apart targets with different expression levels, since stronger bonds lead to appreciable adsorption for a broader range of receptors densities, as proved in \cite{fran-pnas}. In practice, ligands optimization is complicated by these contrasting requirements. 
%

The problems previously described arise from the very nature of multivalent interactions, and thus will affect multivalent constructs regardless of their exact specifics. 
Inspired by previous work on the self-assembly of ligand-receptor coated particles \cite{stefano-nature}, we propose here a functionalization scheme where competition between different receptors is exploited to achieve binding properties which are insensitive to the presence of untargeted receptors, and thus restore selectivity.\\
%
In this ``protected'' scheme, multivalent particles are functionalized with both ligands {\it and} receptors that can bind to them (see Fig.\ref{fig:cases} $d$). From here on, the latter will be called ``protecting'' receptors. %
\begin{figure}[h]
\includegraphics[width=.5\textwidth]{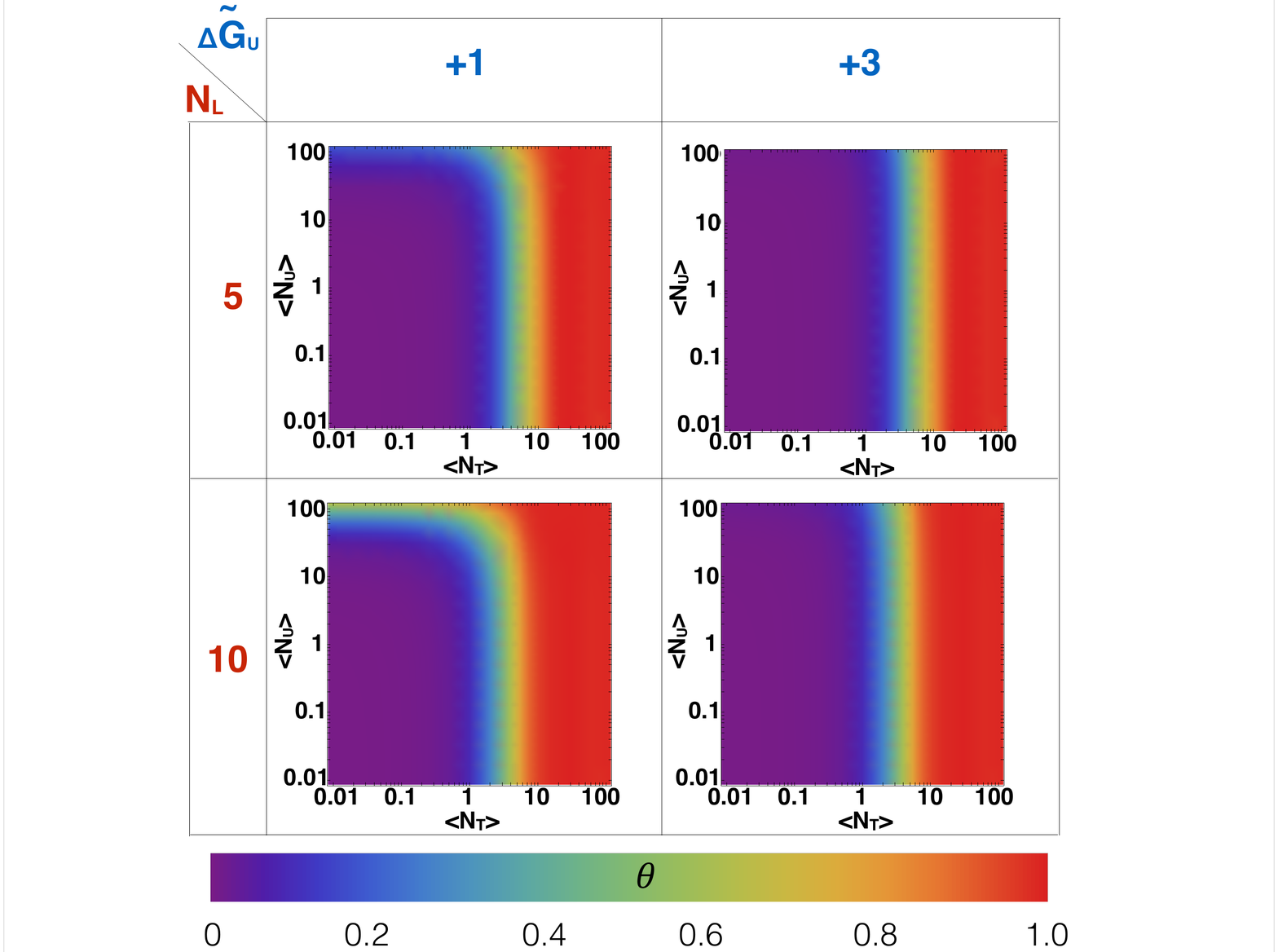}\\
\includegraphics[width=.4\textwidth]{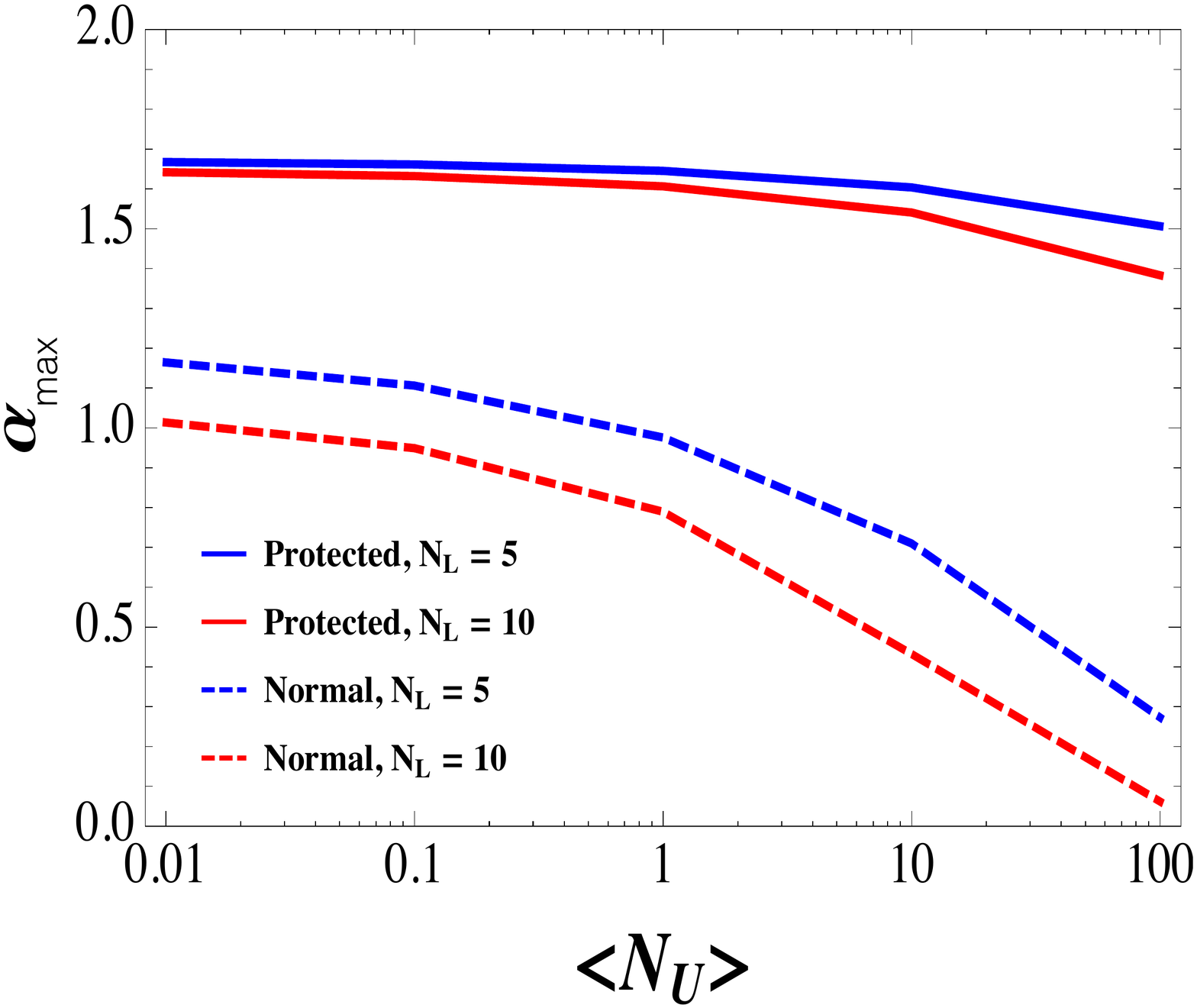}
\caption{Top: As in Fig.~\ref{fig:real-binding} but for protected nanoparticles. Compared to normal particles, non-specific adsorption is strongly reduced. Bottom: Comparison of the maximum selectivity $\alpha_\mathrm{max}$ (Eq.\,\ref{eq:alpha}) between protected and normal nanoparticles, continuous and dashed lines, respectively. Colours denote different number of ligands (5, blue and 10, red). Protected nanoparticles show overall higher selectivity, and display super-selective behaviour, $\mathrm{max}(\alpha) > 1$, in regions where normal particles would perform even worse than monovalent ones. $\fbt=-3$, as in Fig.~\ref{fig:real-binding}, and $\fbin=\fbt$.}
\label{fig:myArchitecture}
\vspace{-.5cm}
\end{figure}
%
%
%
%
%
The effects of such design, reported in Fig.\,\ref{fig:myArchitecture}, are evident comparing with the previous results for normal nanoparticles in Fig.\,\ref{fig:real-binding}. First, the amount of untargeted receptors that protected particles can bear without affecting adsorption is greatly increased, and in fact ideal behaviour is observed for almost all receptors concentrations and binding strengths.
Moreover, super-selective adsorption ( $\alpha > 1$) is recovered in scenarios where the typical multivalent architecture shows sub-optimal behaviour even compared to monovalent particles. Furthermore, no strong deterioration of selectivity with increasing valency is observed for the protected design, whose properties are almost independent on valency.\\
%
%
%
%
Quantifying these effects requires a statistical mechanical description to properly account for the relative weight of all possible binding configurations, but they can be qualitatively understood using a simple microscopic picture.  
Ligands bound to a protecting receptor cannot concurrently bind those on the surface. However, the system is in a dynamic equilibrium where bonds continuously break and form, and on average bonds with surface receptors will occur if their formation is thermodynamically favourable compared to those with protecting receptors. This is typically so only for bonds with the targeted receptors for which ligands are optimised to have a high binding strength, not for others. It is indeed this mechanism that leads to high selectivity for this scheme.
The equilibrium behind it is controlled by the relative energies of these bonds, $\fbt-\fbin$ and $\fbu-\fbin$ ($P$ meaning protecting receptors). It also depends on the number of ligands, protecting, and surface receptors, since their quantity affect the relative entropy of binding. In fact, within a mean field model, one could simply regard the number of ligands or receptors as an additional energy contribution of $\fb_x = - \log N_x$ to be added to the true bond energy. 
The microscopic picture we presented easily describes other trends in this system, such as the increase in selectivity with increasing number of protecting receptors, and with stronger protecting bonds (see the Supplementary Information).
It also explains the increase in the selectivity parameter $\alpha$ for the protected scheme: protecting bonds shift the {\it effective} bond energy without reducing the difference between targeted and untargeted receptors. 
%
In this way, they boosts super-selectivity because they provide a mechanism to make targeted bonds effectively weaker and weaker without however increasing the contribution of non-specific interactions, which would instead come into play with a normal design. In practice, protecting receptors allow to reduce a realistic multi-receptor scenario to an ideal one where only targeted receptors are present.
Before concluding, we discuss a possible practical realisation of this ``protected'' architecture. Although this scheme is very general and various solutions can be envisaged, we would like to focus on a specific system which we deem very promising, colloids coated with DNA (DNACCs). 
Protected DNACCs have already been experimentally implemented to study self-assembly \cite{mirjam-self,lorenzo-pnas1}. This system, where DNA acts both as ligand and receptor, represents in itself a very interesting and powerful solution. DNACCs are already under intense investigation in drug-delivery and biosensing \cite{mirkin}, but even more importantly DNA can bind to a large variety of molecules. Short DNA strands, so-called ``aptamers'', are widely used as targeting ligands for proteins, and a wide literature exists regarding their optimization \cite{review-translational}. 
Such knowledge combined with extensively validated models to calculate DNA-DNA binding energies \cite{santalucia,nupack-algo} can be used as a general and powerful platform for the design of protected targeting constructs. 
One simple way to implement the proposed scheme would be to functionalize nanoparticles with both the targeting aptamer and its complementary DNA sequence, which would act as its protecting receptor. DNA would also provide an easy way to tune the strength of protecting bonds, by simply changing the nucleotides sequence of protecting strands.\\

In conclusion, in this Letter we addressed selective targeting using multivalent nanoparticles via a statistical mechanical model. This was done by combining previous results \cite{fran-pnas,stefano-jcp} in a unified theory, allowing to unravel the physics behind non-specific binding arising from the presence of multiple receptor types.
We showed how multivalent particles are more sensitive to non-specific binding compared to monovalent ones, and that their selectivity rapidly deteriorates in presence of different receptors types.
To solve this problem, we proposed and justified a simple design principle: Beside targeting ligands, nanoparticles should also be coated with protecting receptors able to bind them, shielding their interactions with untargeted receptors. We showed that this mechanism prevents non-specific adsorption, and increases selectivity under conditions where typical nanoparticles would not work. Protecting receptors have been previously shown to enhance the self-assembly kinetics of DNACCs \cite{stefano-nature}, a very different problem from the one treated here.
They could also similarly prove beneficial for other soft-matter systems, e.g. in patchy particles on substrates when competing interactions are present \cite{patchy}. We finally stress that, in the context of nanomedical applications,
the problem and the solution proposed here do not apply only to nanoparticles, or to cells targeting. Instead, they emerge whenever multivalent constructs are used, e.g. in functionalized polymers or surfaces that recognize specific biomarkers via multiple binding sites.\\
%
%
%
%

I would like to acknowledge Daan Frenkel, Tine Curk, Bortolo Mognetti, Jure Dobnikar and Nicolas Tito for reading the manuscript and discussing multivalent targeting, and all members of the GCS group of the University of Milano-Bicocca for the feedback. I also acknowledge the BAIC-SME in Beijing for financial support.




\end{document}